\documentclass[referee]{aa}
\usepackage{graphicx}
\def\logz{\lbrack\hbox{Fe/H}\rbrack}
\begin{document}

\title{WFPC2 Observations of Two dSph Galaxies in the M81 Group
\thanks{Based on observations made with the NASA/ESA Hubble Space
Telescope.  The Space Telescope Science Institute is operated by the
Association of Universities for Research in Astronomy, Inc. under NASA
contract NAS 5--26555. Based in part on observations obtained with the
Apache Point Observatory 3.5-meter telescope, which is owned and operated
by the Astrophysical Research Consortium.}}
\titlerunning{Dwarf spheroidal galaxies in the M81 group}
\author{I.D.Karachentsev \inst{1} \and M.E.Sharina \inst{1,10}
\and A.E.Dolphin \inst{2}
\and D.Geisler \inst{3}
\and E.K.Grebel \inst{4,6}\thanks{Hubble Fellow}
\and P.Guhathakurta \inst{5}\thanks{Alfred P.\ Sloan Research Fellow}
\and P.W.Hodge \inst{6}
\and V.E.Karachentseva \inst{7}
\and A.Sarajedini \inst{8}
\and P.Seitzer \inst{9}}
\authorrunning{I.D.Karachentsev et al.}
\institute{Special Astrophysical Observatory, Russian Academy
of Sciences, N.Arkhyz, KChR, 369167, Russia,
\and Kitt Peak National Observatory, National Optical Astronomy Observatories,
P.O. Box 26732, Tucson, AZ 85726, USA
\and Departamento de Fisica, Grupo de Astronomia, Universidad de Concepcion,
Casilla 160-C, Concepcion, Chile
\and Max-Planck-Institut f\"{u}r Astronomie, K\"{o}nigstuhl 17, D-69117
Heidelberg, Germany
\and UCO/Lick Observatory, University of California at Santa Cruz, Santa Cruz,
CA 95064, USA
\and Department of Astronomy, University of Washington, Box 351580, Seattle, WA
98195, USA
\and Astronomical observatory of Kiev University, 04053, Observatorna 3, Kiev,
Ukraine
\and Astronomy Department, Wesleyan University, Middletown, CT 06459, USA
\and Department of Astronomy, University of Michigan, 830 Dennison Building,
Ann Arbor, MI 48109, USA
\and Isaac Newton Institute, Chile, SAO Branch}
\date{Received:  November 2000}

\abstract{
We have obtained HST WFPC2 and ground-based images of two low surface
brightness dwarf spheroidal galaxies in the M81 group, FM1 and
KKH57. Their colour-magnitude diagrams show red giant branches with tips
at $I = 23.77 \pm 0.14$ and $I = 23.97 \pm0.17$, respectively. The derived
true distance moduli, $27.66 \pm 0.16$ and $27.96 \pm 0.19$, agree well
with the mean distance modulus
of the M81 group, $27.84 \pm 0.05$. Absolute $V$ magnitudes
of the galaxies ($-11.46$ and $-10.85$), their colours ($(B-V) = 0.88$ and
0.80), and central surface brightnesses ($\Sigma_{0,V}$  = 24.8 and 24.4
mag/$\sq\arcsec$) are in the range of other dSph companions of
M81, M31, and Milky Way. With two new objects the maximum projected
radius of the dwarf spheroidal subsystem around M81 is 380 kpc.
}
\maketitle

\section{Introduction}
The first three dwarf spheroidal galaxies 
in the M 81 group, DDO~44, DDO~71, and
DDO~78, were discovered more than 40 years ago by van den Bergh (1959).
In the following we call dwarf galaxies dwarf spheroidals (dSphs) 
if they are diffuse
spheroidal objects with absolute $V$ magnitudes of M$_V \ga -14^{m}$,
surface brightnesses of $\Sigma_V \ga 22$ mag arcsec$^{-2}$,
and H\,{\sc i} masses of $M_{\rm HI} \la 10^5 M_{\odot}$ (Grebel 2000).
More recent studies (Karachentseva 1968, B\"{o}rngen \& Karachentseva 1982,
Karachentsev 1994 and Caldwell et al. 1998) have found six more
dSphs around M81 -- K61, K64, BK5N, BK6N, KK77, and F8D1.
All nine of these objects have been resolved into stars in Wide Field
Planetary Camera 2 (WFPC2) images and confirmed as members of the M81 group
(Caldwell et al. 1998, Karachentsev et al. 1999, Karachentsev et al. 2000).
In this paper we report observations with WFPC2 of two more dSph
galaxies in the M81 group: FM1 and KKH57. The former object was
discovered by Froebrich \& Meusinger (2000) on digitally stacked Schmidt
telescope plates. The latter was found on POSS-II film copies by
Karachentsev et al. (2001). Both galaxies are resolved into stars,
yielding distances that are consistent with their membership in the M81
group. As a result, the present number of known dSph systems
in the M81 group has reached eleven, being comparable with the number of
known dSph objects in the Local Group (van den Bergh 2000).

\section{Ground-based Observations}

$B,V$ CCD images of FM1 were obtained by E.Grebel and I.Karachentsev on
February 3, 2000 using the 3.5m APO telescope.  Part of the $V$ image,
which was taken with $1\farcs5$ seeing and a 600s exposure, is shown
in Figure 1. The galaxy is very diffuse, with a bright star to the
northwest making photometry difficult. To the southwest is a background
galaxy. Integrated photometry of FM1 has been performed by L.Makarova
with increasing circular apertures. The sky level has been approximated
by a two-dimensional polynomial, using regions with few stars near the edges
of the images. The galaxy magnitude in each band has then been measured
as the asymptotic value of the derived growth curve. The results are
$B_T = 17.50 \pm 0.15$ and $V_T = 16.62 \pm 0.15$, values consistent
with the photographic magnitudes $B_T = 17.2$ and $R_T = 15.9$ measured
by Froebrich \& Meusinger (2000).

KKH57 was observed in $B$ and $V$ with the APO telescope, as well as 
in $V$ and $I$ with the 6m SAO (Russia) telescope on December 9, 1999.
Figure 2 shows a 1200s $V$ image from the 6m telescope.  The galaxy looks
more compact than FM1 and granulated because of the presence of stars
near the detection threshold. Surface photometry of KKH57 obtained in
the same manner as described above yields
integrated magnitudes $B_T = 17.86 \pm 0.15$ and $V_T = 17.06 \pm 0.15$.

\section{WFPC2 Observations}

\textit{Hubble Space Telescope} WFPC2 observations of FM1 and KKH57 were
obtained on July 8 and September 5, 2000, respectively, as part of the
HST snapshot survey of probable nearby galaxies (program GO 8601, PI:
P.Seitzer).  Each galaxy was imaged in F606W and F814W with exposure
times of 600s each, with the galaxy centres located on the WFC3 chip.
Figures 3 and 4 show the galaxy images resulting from the contribution of
both filters to remove cosmic rays.

The photometric pipeline used for the snapshot survey has been described
in detail in Dolphin et al. (2001), and what follows is only a summary.
After obtaining the calibrated images from STScI, cosmic ray cleaning
was made with the HSTphot (Dolphin 2000a) \textit{cleansep} routine,
which cleans images taken with different filters by allowing for a
colour variation.  Stellar photometry was then obtained with the HSTphot
\textit{multiphot} routine, which measures magnitudes simultaneously in
the two images, accounting for image alignment, WFPC2's wavelength-dependent
plate scale, and geometric distortion.
The final photometry was then made using aperture corrections to
a $0\farcs5$ radius, and the Dolphin (2000b) charge-transfer inefficiency
correction and calibration applied. We estimate the aperture corrections
in the three wide field chips to be accurate to 0.05 magnitudes. Because
of the small sky coverage of the planetary camera (PC), and lack of stars for
an accurate aperture correction, the PC photometry is omitted from our
results. Aditionally, stars with signal-to-noise $< 5$, $ \vert chi \vert > 2.0 $, or
$ \vert sharpness \vert > 0.4 $ in either exposure were eliminated from the final
photometry list, in order to minimize the number of false detections.
 Finally, the F606W and F814W instrumental magnitudes were converted
to the standard $V$, $I$ system folowing the "synthetic" transformations of
Holtzman et al. (1995). We used parameters of transformation from their
Table 10 taking into account different relations for blue and red stars
separately. Because we used the non-standard $V$ filter F606W instead of
F555W, the resulting $I$ and especially $V$ magnitudes may contain systematic
errors. However, when comparing our {F606W, F814W} photometry of other
snapshot targets with ground-based $V$,$I$ photometry we find that the
transformation uncertainties, $\sigma_ {I}$ and  $\sigma_ {V - I}$, are
within $0.^{m}05$ for stars with colors of $0 < (V-I) < 2$.
 The resulting $(V-I)$, $I$
colour-magnitude diagrams (CMDs) are shown in Figures 5 and 6.

We also measured integrated properties of the two galaxies using aperture
photometry in circular apertures, finding $V_T = 16.60 \pm 0.15$ and
$I_T = 15.21 \pm 0.15$ for FM1 and $V_T = 17.00 \pm 0.15$ and
$I_T = 15.82 \pm 0.15$ for KKH57, as well as the central
surface brightness and the exponential scale length presented in Table 1.

\section{Colour-magnitude Diagrams and Distances}

In Figures 5 and 6 the left panels show the CMDs for the central WFC3 fields
covering the main galaxy bodies. The middle panels represent the CMD for
the neighbouring halves of WFC2 and WFC4 ($x < 425$ in WFC2 and $y < 425$
in WFC4), and the right panels are
comprised of stars found in the remaining outer halves of WFC2 and WFC4.
As seen from these CMDs, the observed stellar populations of both galaxies
consist predominately of red stars.
To measure the distances to these galaxies, we apply the red giant branch
(RGB) tip method described by Lee et al. (1993). The Gaussian-smoothed
$I$-band luminosity functions for the central fields of FM1 and KKH57
are shown in the top panels of Figure 7. Only red stars with colors
$1 < (V - I) < 2$ were considered. Measurment the RGB tip positions
can be done using a Sobel filter. Following Sakai et al. (1996), we
apply an edge-detection filter, which is a modified version of a Sobel
kernel ([-1,0,+1]), to the luminosity functions to determine objectively
the positions of the TRGB. The results of convolution are shown in the
bottom panels of Figure 7. The positions of the TRGB are identified with
the highest peak in the filter output function. For FM1 and KKH57 we
obtain $I_{TRGB} = 23.77 \pm 0.14$ and $23.97 \pm 0.16 $, respectively.
  Artificial star tests were also made, so that the accuracy and completeness
of the crowded WF3 photometry could be measured. We can conclude from
these simulations that the detection rate dropped to 50\% at $I_{lim} = 24.8 $
 and $ V_{lim} = 25.9 $. The scatter of colors and magnitudes for the detected
stars increases towards faint magnitudes. At the level of $ I \sim 23.^{m}8 $
the position of the TRGB shifts to a brighter ($0.^{m}06 $) and bluer
($ 0.^{m}04 $) magnitude due to stellar crowding. The same effect has been
shown by Madore \& Freedman (1995).

Adopting an RGB tip absolute magnitude of
$M_I = -4.03 \pm 0.05$ estimated from the theoretical models of Girardi
et al. (2000) and the semi-empirical calibration of Lee et al. (1993)
and Galactic extinction values of $A_I = 0.14$ (FM1) and $A_I = 0.04$
(KKH57), calculated from the maps of Schlegel et al. (1998), we derive
distance moduli of $\mu_0 = 27.66 \pm 0.17$ and $\mu_0 = 27.96 \pm 0.19$,
respectively. The quoted errors include the error in the TRGB detection,
and also uncertainties of the HST photometry
zero point ($ \sim 0.^{m}05 $), the aperture corrections ($\sim 0.^{m}05$),
and crowding effects ($ \sim 0.^{m}06 $) added in quadrature.

Three solid lines in the left panels of Figures 5 and 6 are globular
cluster fiducials from Da Costa \& Armandroff (1990), adjusted for the
reddening and distance of each galaxy. The fiducials cover the range of
metallicity values $\logz = -2.17$ dex for M15, $-1.58$ dex for M2, and
$-1.29$ dex for NGC 1851 (from left to right).

\section{Integrated Properties}

Table 1 presents a summary of basic properties of FM1 and KKH57 in the
same format as for eight previously-considered dSphs in the M81 group
(Karachentsev et al. 2000). Its lines contain: (1,2) equatorial coordinates,
(3--5) Galactic extinction from Schlegel et al.(1998), (6) galaxy dimensions
along the major and minor axes approximately corresponding to a level of
$B$ = 26.5 mag/$\sq\arcsec$, (7) integrated $V$ magnitude, (8,9) integrated
colours, (10) observed central surface brightness, (11) exponential scale
length in arcsec averaged over the $V$ and $I$ bands, (12) apparent $I$
magnitude of the RGB tip, (13) true distance modulus, (14,15) mean
reddening-corrected colour of the RGB tip measured at an absolute
magnitude $M_I = -3.5$, as recommended by Lee et al. (1993) and
corresponding mean metallicity, (16) mean metallicity on the scale of
Carretta \& Gratton (1997), (where the errors were calculated by carying
the color uncertainty into the Lee et al. formula, as well as the intrinsic
uncertainty in such a calibration, and the difference between the metallicity
scales is discussed at length by Carretta \& Gratton, 1997), (17,18) linear
diameter and total absolute magnitude of the galaxy with the mean
distance modulus of 27.84, (19,20) angular and linear
projected separation of the galaxy from M81, (21) morphological type, (22)
number of globular cluster candidates in each galaxy.

\section{Discussion}

Froebrich \& Meusinger (2000) imaged FM1 in the H$_{\alpha}$ line, finding
no indication of significant H$_{\alpha}$ emission. Karachentsev et al.
(2001) observed KKH57 in the 21cm line with the 100-m Effelsberg telescope
and obtained an upper limit of the H\,{\sc i} flux of 5 mJy, corresponding to an
H\,{\sc i} mass limit of $4\cdot10^5$ solar masses. This limit is similar to
the upper H\,{\sc i} mass limit found for Local Group dSphs.
The symmetric shape of
both galaxies, their smooth surface brightness profiles, the reddish
integrated colours, the lack of an appreciable amount of hydrogen, and the
absence of a population of bright blue stars (see Figures 5 and 6) favour
the classification of FM1 and KKH57 as dSph galaxies.

According to the summary of distance moduli for 11 members of the M81
group (Table 4 in Karachentsev et al. 2000) measured via the RGB tip or
Cepheids, the mean distance modulus of the group is $27.84 \pm 0.05$.
Therefore, the distance moduli of these two new dSph galaxies,
$27.66 \pm 0.17$ and $27.96 \pm 0.19$, are consistent with their
membership in that group.

Judging by the absolute $V$-band magnitudes ($-11.46$ and $-10.85$) and linear
diameters (0.97 kpc and 0.64 kpc), FM1 and KKH57 are among the faintest
known dSph members of the M81 group. Nevertheless, their luminosities
exceed by about two magnitudes the luminosity of the faintest dSphs in the
Local Group: Draco ($-8.6$), Ursa Minor ($-8.9$), and Andromeda V ($-9.1$).
It is quite likely that similar systems exist in the M81 group as well,
but have not yet been detected. (Two new extremely low surface brightness
objects of the dSph type were found in the M81 group by Karachentseva and
Karachentsev in December 2000). The presence of Galactic cirrus in that
direction (Sandage 1976) makes such a search extremely difficult.
The Local Group contains 10 dSphs with absolute magnitudes brighter than
$M_V = -10^{m}$, very similar in number to the 11 dSphs currently known
down to this magnitude limit in the M81 group.  The Local Group contains
at least seven dSphs fainter than $M_V = -10^{m}$, which lets
near-infrared searches for faint extended objects in the M81 group appear
a promising approach to detect additional, faint dwarf members.
 
Figure 8 shows the positions of ten dSph companions of M81 in equatorial
coordinates. The most remote companion, KKH57, has a projected distance
of 381 kpc from M81. The distribution of dSphs looks to be asymmetric,
which can be caused by the presence of Galactic cirrus in the M81 group
direction.

We searched for globular clusters in FM1 and KKH57 but found no candidates
with the appropriate range of colours, magnitudes and half-light radii
defined by Milky Way globulars. This null result is not surprising, given
the low expected value of the specific frequency of globular clusters in
low-luminosity dSphs (Miller et al. 1998) -- the presence of even one
globular in FM1 (the brighter of the two) would produce a specific
frequency $S_N$ of nearly 30, much larger than any galaxy in the Miller
et al. (1998) sample.

\acknowledgements
{ We are grateful to Lidia Makarova for providing us with surface photometry
of the galaxies.
Support for this work was provided by NASA through grant GO--08192.97A from
the Space Telescope Science Institute, which is operated by the Association
of Universities for Research in Astronomy, Inc., under NASA contract
NAS5--26555.  IDK, VEK, and EKG acknowledge partial support through the
Henri Chr\'{e}tien International Research Grant administered by the American
Astronomical Society.  EKG acknowledges support by NASA through grant
HF--01108.01--98A from the Space Telescope Science Institute. This work
has been partially supported by the DFG--RFBR grant 98--02--04095.}

\onecolumn
\begin{table}
\caption{Properties of two new dSph galaxies in the M81 group}
\begin{tabular}{lcc} \\  \hline
 Parameter                &       FM1        &      KKH57       \\
\hline
 RA (2000.0)              &  09$^h$45$^m$10$\fs$0 &    10$^h$00$^m$16$\fs$0 \\
 Dec (2000.0)             & +68$\degr$45$\arcmin$54$\arcsec$& +63$\degr$11$\arcmin$06$\arcsec$ \\
 $E(B-V)$                 &      $0.08$      &      $0.02$      \\
 $A_V$                    &      $0.24$      &      $0.07$      \\
 $A_I$                    &      $0.14$      &      $0.04$      \\
 $a\times b,\; (\arcmin)$ & $0.9\times0.8$   & $0.6\times0.5$   \\
 $V_T$                    &  $16.62 \pm 0.15$ & $17.06 \pm 0.15$ \\
 $(B-V)_T$                &  $0.88 \pm 0.15$ & $0.80 \pm 0.15$  \\
 $(V-I)_T$                &  $1.39 \pm 0.15$ & $1.18 \pm 0.15$  \\
 $\Sigma_{0,V}$           &  $24.8 \pm 0.2$  & $24.4 \pm 0.2$   \\
 $h,\; (\arcsec$)         &     $21$         &     $14$         \\
 $I$(TRGB)                & $23.77 \pm 0.14$ & $23.97 \pm 0.16$ \\
 $\mu_0$                  & $27.66 \pm 0.17$ & $27.96 \pm 0.19$ \\
 $(V-I)_{0,-3.5}$         & $1.37 \pm 0.05$  & $1.42 \pm 0.05$  \\
 $\logz$                  & $-1.6 \pm 0.6$   & $-1.4 \pm 0.6$   \\
 $\logz$ (CG97)           & $-1.3 \pm 0.6$   & $-1.2 \pm 0.6$   \\
 $D_{26.5} ,\;$ kpc       &      $0.97$      &      $0.64$      \\
 $M_V$                    &    $-11.46$      &    $-10.85$      \\
 $r$(M81),$\; (\arcmin)$  &      59          &     354          \\
 $R$(M81),$\;$ kpc        &      64          &     381          \\
 Type                     &     dSph         &    dSph          \\
 $N_{gc}$                 &       0          &      0           \\
\hline
\end{tabular}
\end{table}

\begin{figure}[hbt]
\vbox{\includegraphics{MS10508f1.ps}}\par
\vspace{17cm}
\caption{$V$-band image of FM1 obtained with the 3.5m APO telescope. The horizontal line indicates 30''. North is up, and the east is to to the left.}
\end{figure}
\begin{figure}[hbt]
\vbox{\includegraphics{MS10508f2.ps}}\par
\vspace{17cm}
\caption{$V$-band image of KKH57 obtained with the 6m SAO telescope. The arrows point to the north and east.}
\end{figure}
\begin{figure}[hbt]
\vbox{\includegraphics{MS10508f3.ps}}\par
\vspace{17cm}
\caption{WFPC2 image of FM1 produced by combining the two 600s exposures taken through the F606W and F814W filters. The galaxy is centred in the WFC3 chip (WF3-FIX mode). North is up and east is to the left.}
\end{figure}
\begin{figure}[hbt]
\vbox{\includegraphics{MS10508f4.ps}}\par
\vspace{17cm}
\caption{WFPC2 image of KKH57 produced by combining the two 600s exposures taken through the F606W and F814W filters. The galaxy is centred in the WFC3 chip. The arrows point to the north and east.}
\end{figure}
\begin{figure}[hbt]
\vbox{\includegraphics{MS10508f5.ps}}\par
\vspace{20cm}
\caption{WFPC2 colour-magnitude diagram for the dSph galaxy FM1. The left panel shows stars in WFC3 (the center of the galaxy), the middle panel the ``medium'' field (neighbouring halves of the WFC2 and WFC4 chips), and the right panel the ``outer'' field (remaining halves of the WFC2 and WFC4 chips). Each of these three fields covers an equal area of $750 \times 750$ pixels. The solid lines in the left panel show the red giant branches of globular clusters with different metallicities: M15 ($-2.17$ dex). M2 ($-1.58$ dex), and NGC~1851 ($-1.29$ dex).}
\end{figure}
\begin{figure}[hbt]
\vbox{\includegraphics{MS10508f6.ps}}\par
\vspace{20cm}
\caption{WFPC2 colour-magnitude diagram for the dSph galaxy KKH57, displayed as in Figure 5.}
\end{figure}
\begin{figure}
\vbox{\includegraphics{MS10508f7a.ps}}\par
\vspace{8cm}
\vbox{\includegraphics{MS10508f7b.ps}}\par
\vspace{15cm}
\caption{The Gaussian-smoothed I-band luminosity function restricted to red
stars with colors between $ 1 < (V - I) < 2 $ (top), and output of an
edge-detection filter applied to the luminosity function (bottom) for FM1
and KKH57. The position of the TRGB is indicated by the highest peak in
the output functions.}
\end{figure}
\begin{figure}[hbt]
\vbox{\includegraphics{MS10508f8.ps}}\par
\vspace{20cm}
\caption{The distribution of dSph galaxies in equatorial coordinates around M81. The four bright galaxies are indicated with large filled circles; dwarf spheroidals are indicated with small open circles.
The asymmetric distribution of dSphs can be caused by the presence of
Galactic cirrus in the M81 group direction.}
\end{figure}

\begin{thebibliography}{}

\bibitem[1982]{b3}B\"{o}rngen F., Karachentseva V.E. 1982, Astron. Nachr. 303, 189
\bibitem[1998]{b4}Caldwell N., Armandroff T.E., Da Costa G.S., Seitzer P., 1998, AJ 115, 535
\bibitem[1997]{b23}Carretta E., Gratton R.G., 1997, A\&AS 121, 95
\bibitem[1990]{b5}Da Costa G.S., Armandroff T.E., 1990, AJ 100, 162
\bibitem[2000]{b19}Dolphin A.E., 2000a, PASP 112, 1383
\bibitem[2000]{b20}Dolphin A.E., 2000b, PASP 112, 1397
\bibitem[2001]{b26}Dolphin A.E., Makarova L., Karachentsev I.D., Karachentseva V.E., Geisler D., Grebel E.K., Guhathakurta P., Hodge P.W., Sarajedini A., Seitzer P., 2001, MNRAS in press
\bibitem[2000]{b6}Froebrich D, Meusinger H., 2000, A\&A, A\&AS 145, 229
\bibitem[2000]{b22}Girardi L., Bressan A., Bertelli G., Chiosi C., 2000, A\&AS 141, 371
\bibitem[2000]{b24}Grebel E.K., 2000, 33rd ESLAB Symposium on ``Star Formation from
the Small to the Large Scale'', SP-445, eds.\ F.\ Favata, A.A.\ Kaas, \& A.\
Wilson (Noordwijk: ESA), 87
\bibitem[1995]{}Holtzman J.A., Burrows C.J., Casertano S.,et al, 1995, PASP 107,
  1065
\bibitem[1994]{b8}Karachentsev I.D. 1994, Astron. Astrophys. Trans. 6, 1
\bibitem[1999]{b10}Karachentsev I.D., Sharina M.E., Grebel E.K., Dolphin A.E., Geisler D.,Guhathakurta P., Hodge P.W., Karachentseva V.E., Sarajedini A., Seitzer P., 1999, A\&A 352, 399.
\bibitem[2000]{b11}Karachentsev I.D., Karachentseva V.E.,Dolphin A.E., Geisler D., Grebel E.K., Guhathakurta P., Hodge P.W., Sarajedini A., Seitzer P., Sharina M.E., 2000, A\&A 363, 117
\bibitem[1968]{b12}Karachentseva V.E. 1968, Commun. Byurakan obs. 39, 62
\bibitem[2001]{b13}Karachentsev I.D., Karachentseva V.E, Huchtmeier W.K., 2001, A\&A 366, 428
\bibitem[1993]{b21}Lee M.G., Freedman W.L., Madore B.F., 1993, ApJ 417, 553
\bibitem[1995]{} Madore B.F., Freedman W.L., 1995, AJ 109, 1645
\bibitem[1998]{b14}Miller B.W., Lotz J.M., Ferguson H.C., Stiavelli M., Whitmore B.C., 1998, ApJ 508, L133
\bibitem[1996]{} Sakai S., Madore B.F., Freedman W.L., 1996, ApJ 461, 713
\bibitem[1976]{b15}Sandage A.R., 1976, AJ 81, 964
\bibitem[1998]{b16}Schlegel D.J., Finkbeiner D.P., Davis M., 1998, ApJ 500, 525
\bibitem[1959]{b18}van den Bergh S., 1959, Publ. of D.D.O., v.II, N.5, 147
\bibitem[2000]{b25}van den Bergh S., 2000, The Galaxies of the Local Group,
   Cambridge University Press

\end{thebibliography}
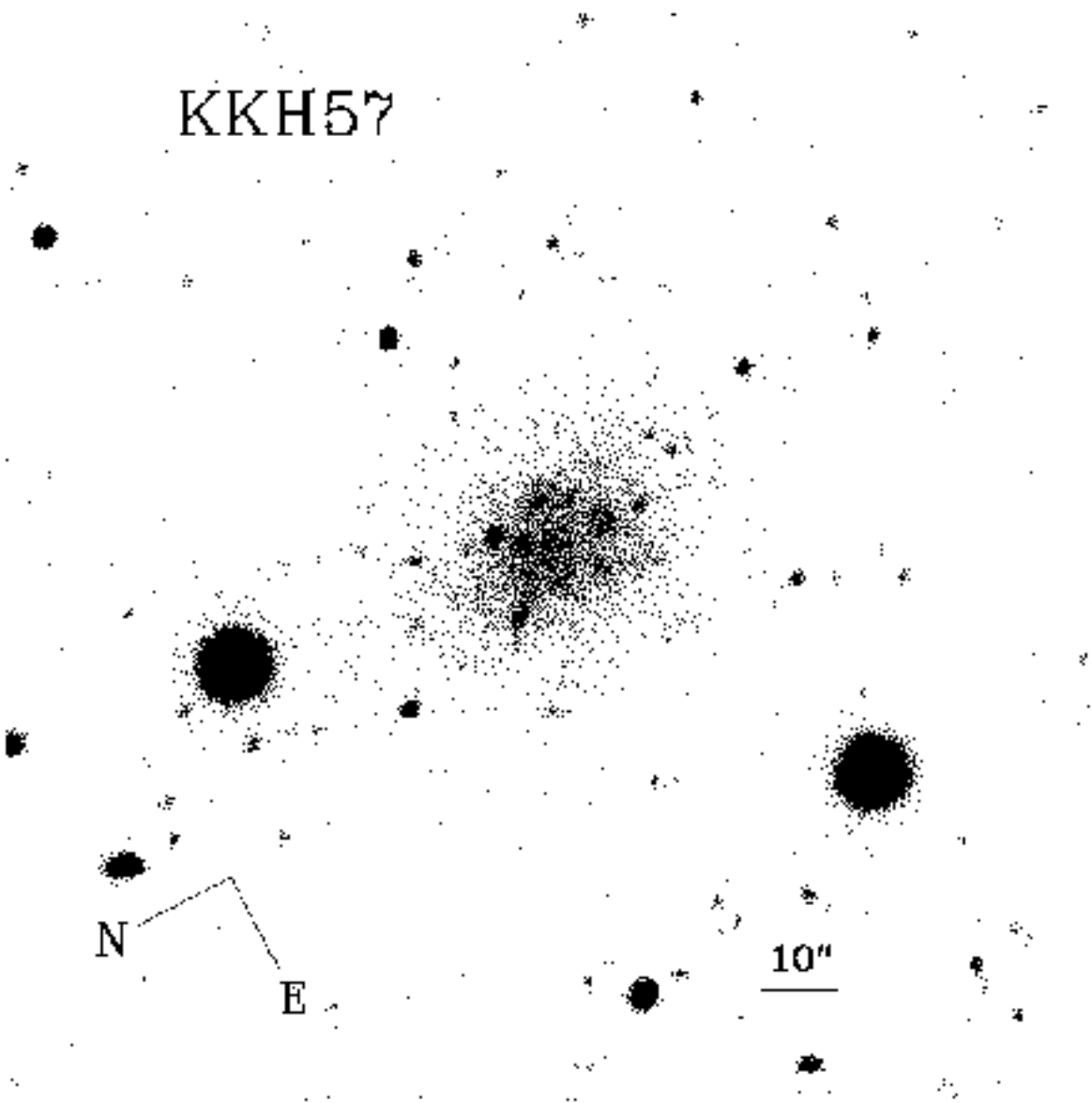
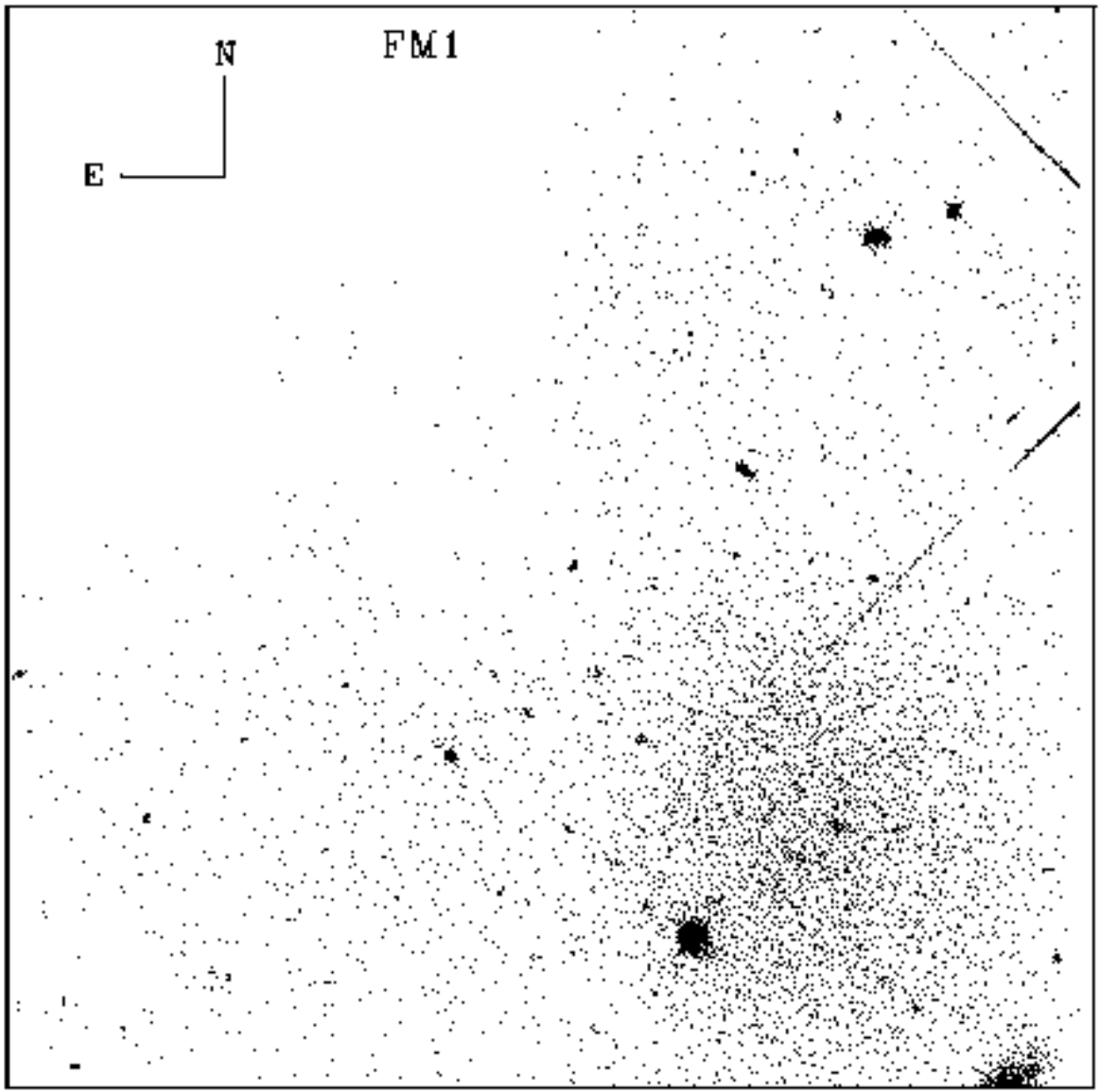
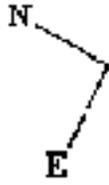
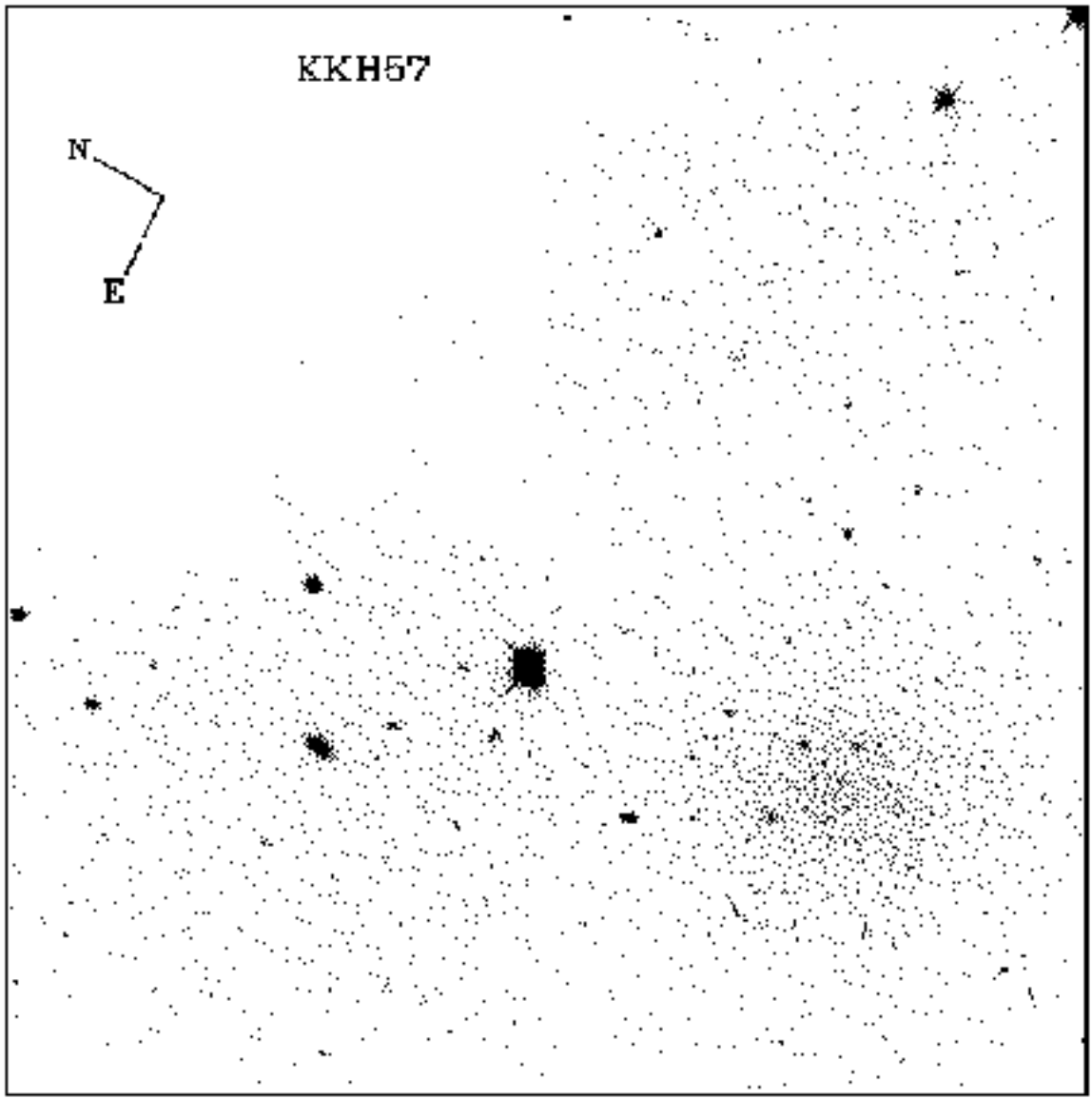
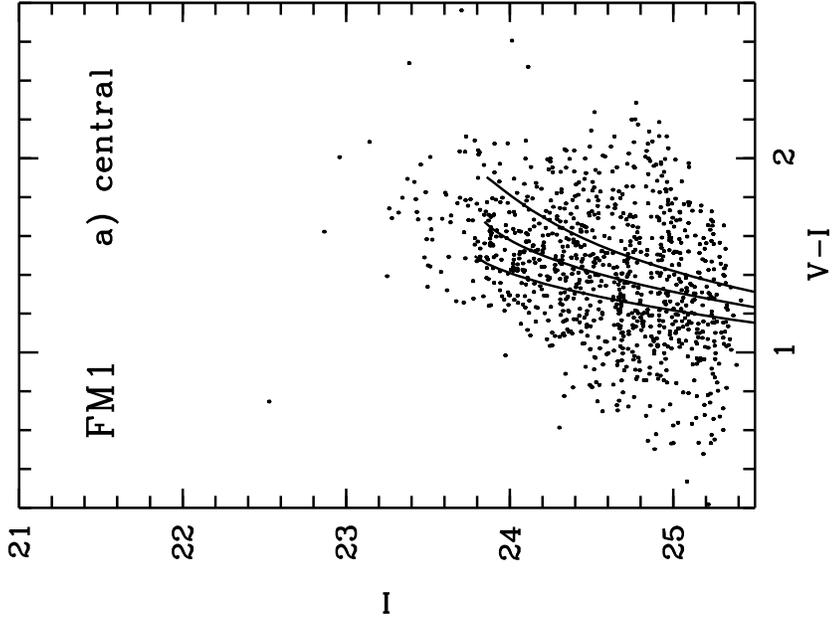
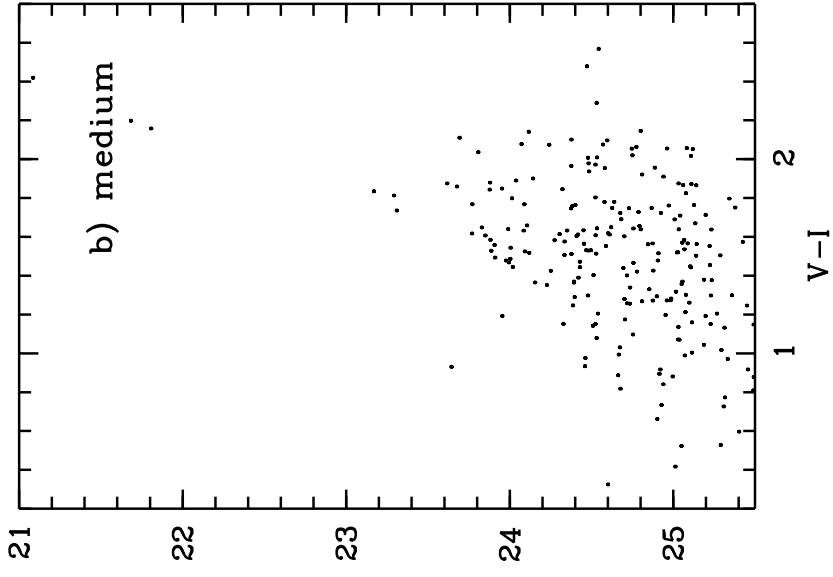
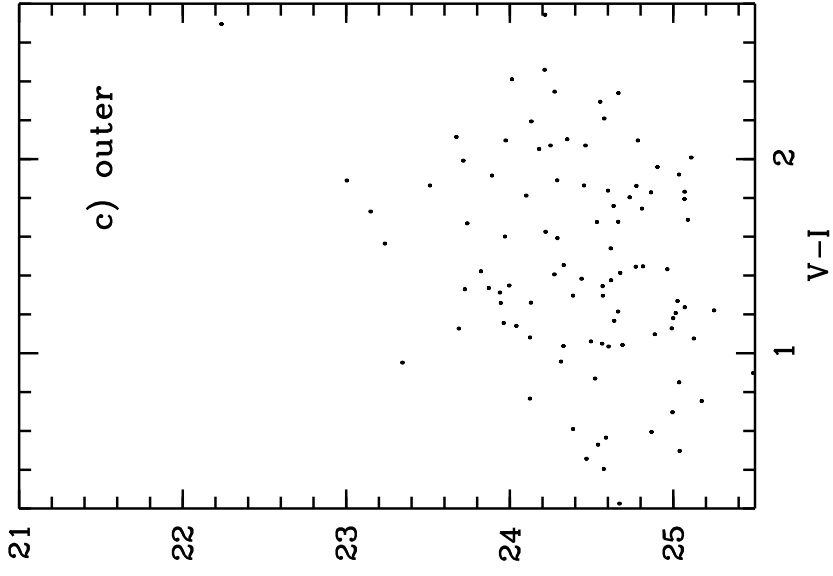
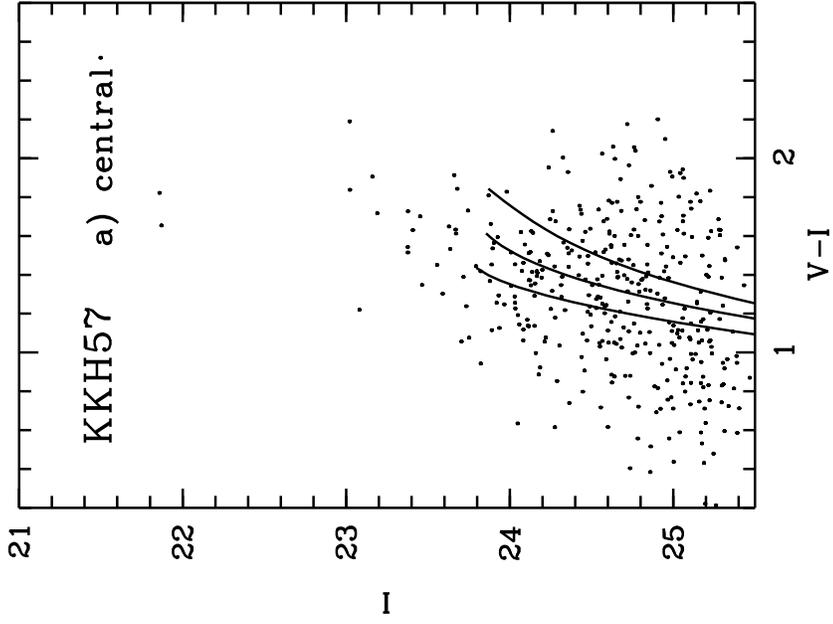
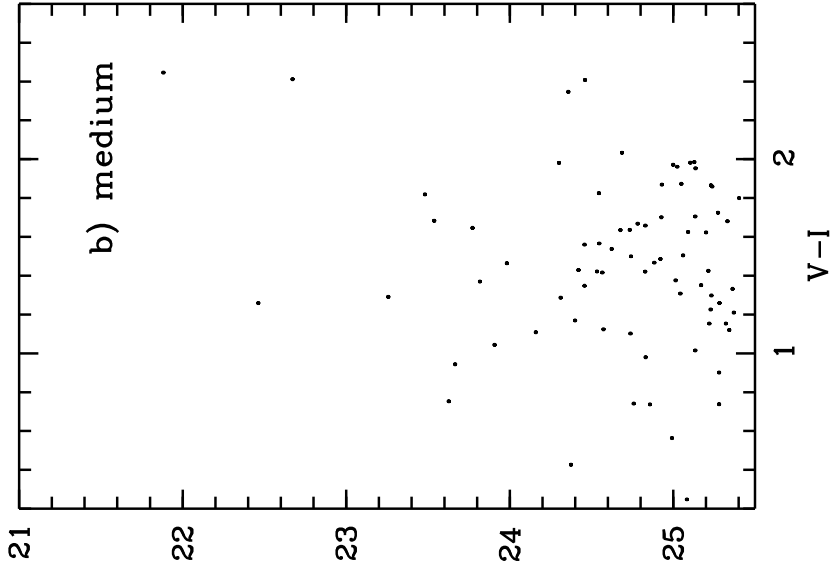
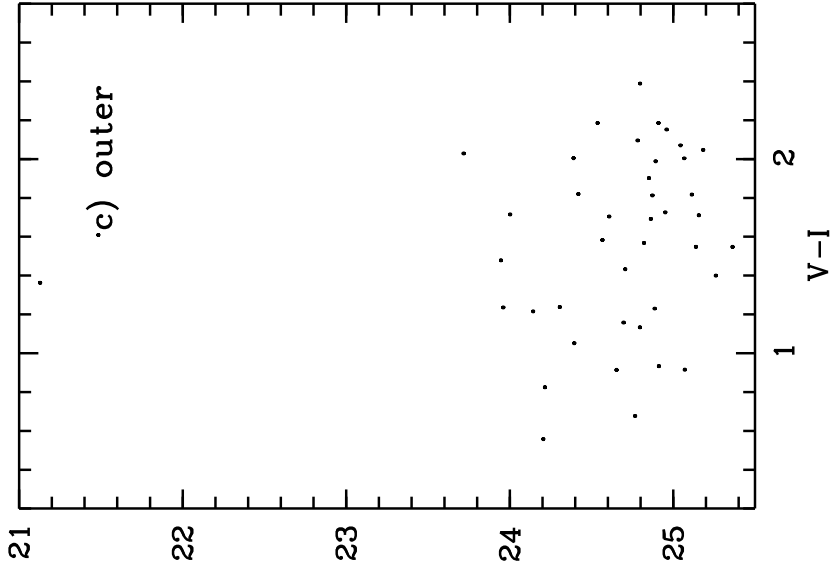
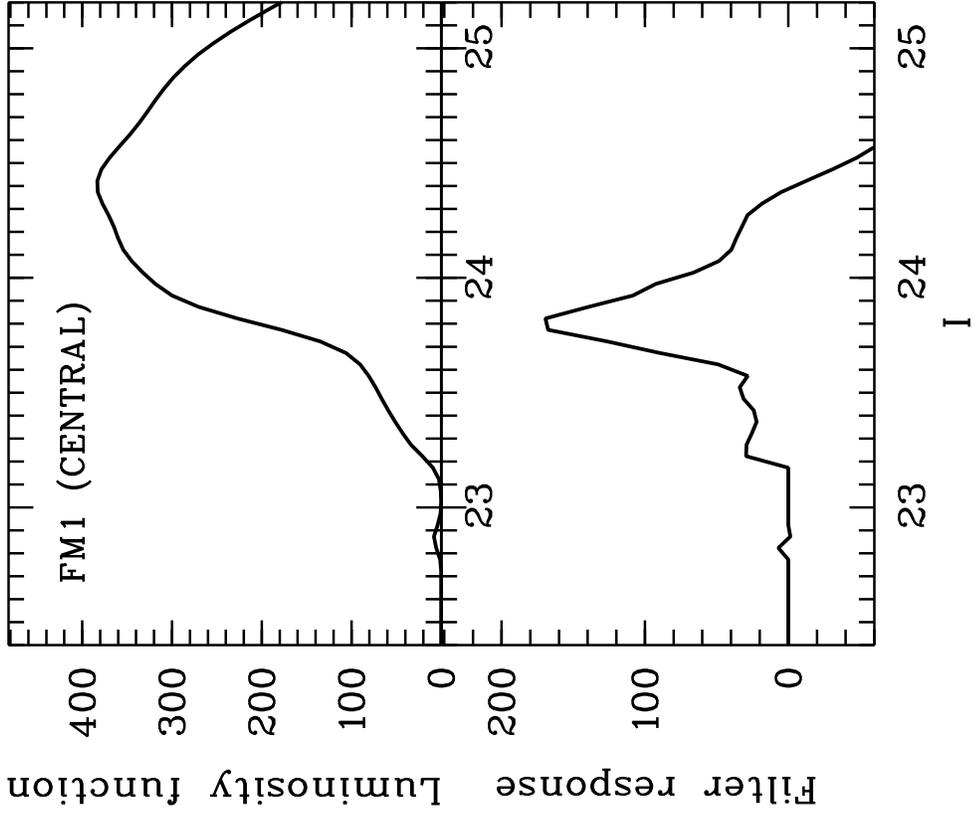
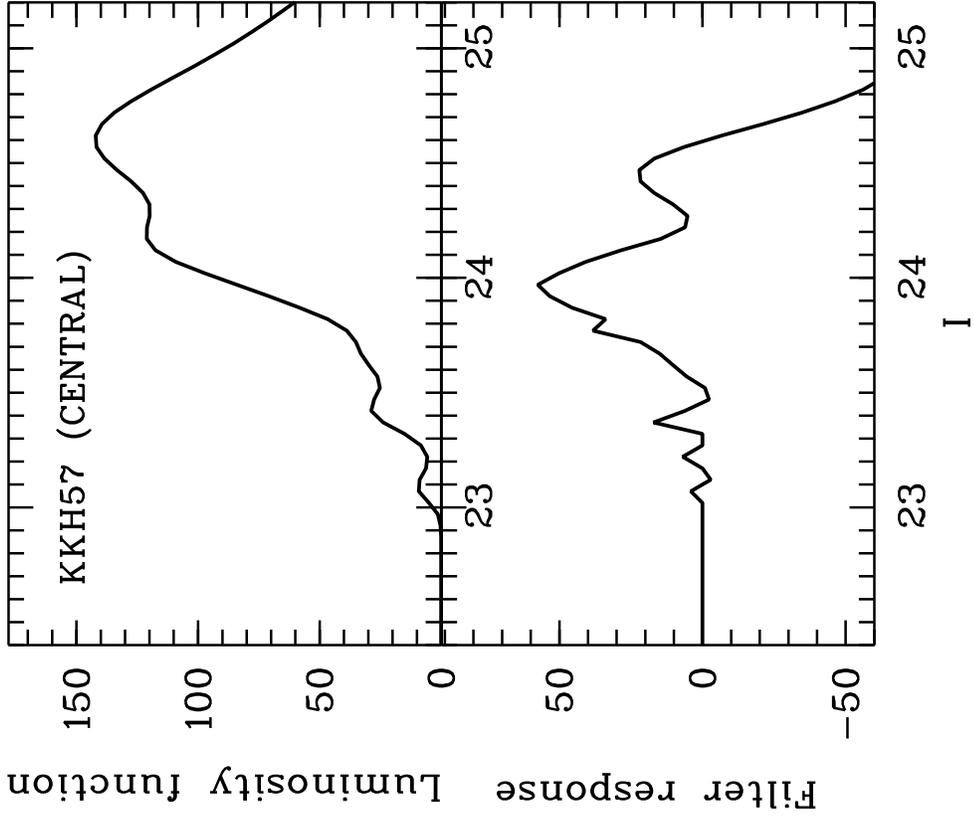
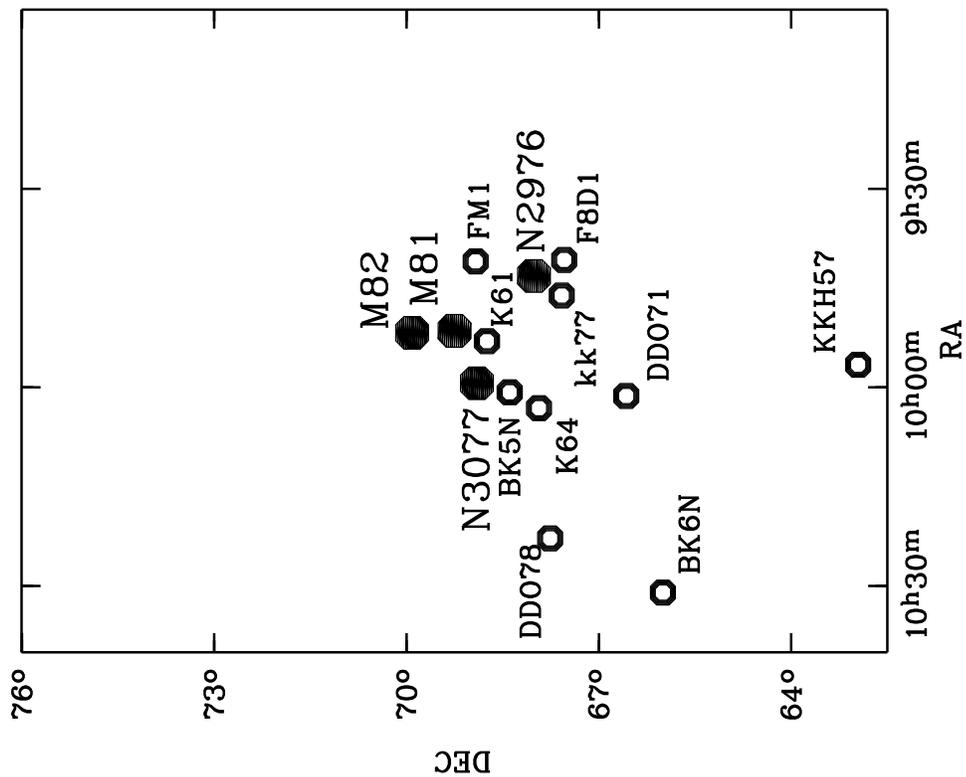
\end{document}